 \documentclass[final,5p,times,twocolumn,authoryear]{elsarticle}

\usepackage{amssymb}
\usepackage{lipsum}

\journal{High Energy Astrophysics}

\begin{document}

\begin{frontmatter}



\title{NuSTAR detection of a broad absorption line in IGR J06074+2205}


\author[first]{Mohammed Tobrej, Binay Rai, Manoj Ghising, Bikash Chandra Paul}
\affiliation[first]{organization={Department of Physics, North Bengal University},
            addressline={ Raja Rammohunpur,  Dist.: Darjeeling}, 
            city={Siliguri},
            postcode={734013}, 
            state={West Bengal},
            country={India}}

\begin{abstract}
We  present the broadband X-ray study of the BeXRB pulsar IGR J06074+2205 using NuSTAR observations. The temporal and spectral characteristics of the source are investigated.  We detect  coherent X-ray pulsations of the source  and determine the spin period evolution. Using the current spin period data of the source, we  show that the source is spinning down at  0.0202(2)\; s\; $day^{-1}$. The pulse profiles are found to evolve with energy and luminosity. Another interesting feature of the source is that the pulse profiles are dual peaked. The dual-peaked pulse profile is characterized by a decreasing secondary peak amplitude with increasing energy. We also observed  a proportionate increase in the primary peak amplitude as the energy increases. The pulse fraction exhibits an overall increasing trend with the energy. The X-ray continuum of the source indicates the existence of a characteristic absorption feature with a centroid energy $\sim 55$ keV,   which may be interpreted as a cyclotron line.  We estimate the corresponding  magnetic field strength to be  $\sim4.74\times10^{12}$ G. A peculiar `10 keV' absorption feature is observed in the X-ray spectra of the second observation.The luminosity measurements predict that the source may be accreting in the sub-critical regime.
\end{abstract}




\end{frontmatter}




\section{Introduction}
\label{introduction}

The source IGR J06074+2205, identified as a Be X-ray pulsar, has been the focus of comprehensive astronomical investigations since its discovery in 2003 through observations facilitated by the International Gamma-Ray Astrophysics Laboratory (INTEGRAL), as documented by \cite{Chenevez}. Subsequent studies, including those by \cite{Reig}, unveiled the pulsar's distinctive 373.2-second pulse period, further showing evidence of its pulsating nature. The nature of the neutron star's companion was initially proposed by \cite{Halpern} and later substantiated by \cite{Tomsick}. The companion star is identified as a B0.5Ve star located at an approximate distance of $\sim$4.5 kpc, as shown by \cite{Reigg}. Observations of the optical companion have provided with intriguing insights to probe  its dynamical behavior. Variability in the $H_{\alpha}$ line, both in terms of line profile and intensity, was observed on monthly time scales in {\it Ref.} \citep{Reigg}. It is predicted by them that there was a reversal from emission to absorption in the $H_{\alpha}$ line, indicating a complete loss of the equatorial disc of the Be star. Recent astronomical events observed by the Advanced Satellite for Cosmology and Astrophysics (ART-XC) on October 03, 2023,   include the detection of an outburst from the pulsar IGR J06074+2205 as reported by \cite{Semena}.  Further observations from the Fermi Gamma-ray Burst Monitor (Fermi-GBM) mission on the same day detected a spin frequency of 2.6700(2) millihertz. Subsequently, \cite{Semena} and \cite{Chenevezz} analyzed the follow-up observations by INTEGRAL, and estimated the source flux in (4-12), (3-10), and (10-25) keV energy bands respectively, providing additional insights into the post-outburst behavior of the source. The most recent observations concerning this source has been reported by \cite{Roy}.

The rich history of  detections of cyclotron lines in X-ray binary pulsars, dating back to the first detection in Her X-1 \citep{Trumper},  paved the way for advancements in understanding the characteristics of pulsars.  In particular, absorption line features within the (5-110) keV energy range are detected in various X-ray binary pulsars, which is documented by \cite{Makishima} and \cite{Grove}. These are known as cyclotron resonance scattering features (CRSFs), which originate from the resonant scattering of hard X-ray photons and electrons in quantized energy levels \citep{Meszaros}. The separation between the quantized energy levels corresponds to the centroid energy of the cyclotron line ($E_{cyc}$), it can be used to  estimate the magnetic field strength of the neutron star. Specifically, the relationship is expressed as $E_{cyc}$ = 11.6 $\times\; B_{12}$ (keV), where $ B_{12}$ denotes the magnetic field expressed in units of $10^{12}$ G. This relationship enables to glean important insights into the physical properties of IGR J06074+2205 and similar pulsars, enhancing our knowledge about the intriguing astrophysical phenomena.

\section{OBSERVATION AND DATA REDUCTION}

The NuSTAR telescope, designed for high-energy X-ray observations, operates within the energy range of 3-79 keV, utilizing two identical X-ray telescope modules in conjunction with separate focal plane detector modules, namely FPMA and FPMB \citep{Harrison}. The detector technology employed in NuSTAR is the Cadmium Zinc Telluride (CdZnTe) pixel detector, known for its efficiency in capturing hard X-rays.
 \begin{table} 
\begin{center}
\begin{tabular}{cccc}
\hline
Observation Date & Observation ID  & Exposure (ks)    &	 
     \\	
\hline
2023-10-08 07:36:09 &90901331002 &		42.7	 \\
2023-10-15 22:36:09 & 90901331004 &	40.7	 \\

\hline
\end{tabular}

\caption{Details of the NuSTAR observations of IGR J06074+2205.Hereafter, we refer these observations as Obs I and Obs II respectively.}  
\label{a}
\end{center}

\end{table} {h}
The observational details for the specific NuSTAR observations are outlined in Table \ref{a}.  The data reduction and analysis are executed here using HEASOFT -v6.32.1, a comprehensive software package designed for high-energy astrophysics research with  CALDB version 20231121.  NUPIPELINE version 0.4.9, a mission-specific software tailored for NuSTAR observations, is then employed for the standard processing and filtering the data. This stage is crucial for enhancing the signal-to-noise ratio and ensuring the reliability of the subsequent analysis. To extract detailed information about the source, such as light curves and spectra, a meticulous approach is considered. A circular extraction region of $90^{\prime \prime}$  centered on the source is chosen. Additionally, another circular region of the same radius is selected away from the source, serving as the background extraction region. This separation aids in isolating the signal from the source against the background noise. The NUPRODUCTS tool is then utilized to generate the light curves and spectra based on the selected extraction regions. This tool streamlines the process of extracting relevant information from the observational data, providing a clear representation of the temporal and spectral characteristics of the source. A correction from the background is applied using the ftool LCMATH to enhance the accuracy of the light curves. This correction mitigates the influence of background noise, ensuring that the observed variations are attributed to the source of interest. Following the background correction, a barycentric correction is implemented using the BARYCORR tool chosing the planetary ephemeris - JPLEPH.430 (refframe-ICRS). Barycentric correction is essential to account for the motion of the Earth in the solar system, aligning the observational times with the solar system barycenter. The correction is crucial for accurate temporal analysis and interpretation of the observed phenomena. 

\section{Timing Analysis}

\begin{figure}

\begin{center}
\includegraphics[angle=0,scale=0.34]{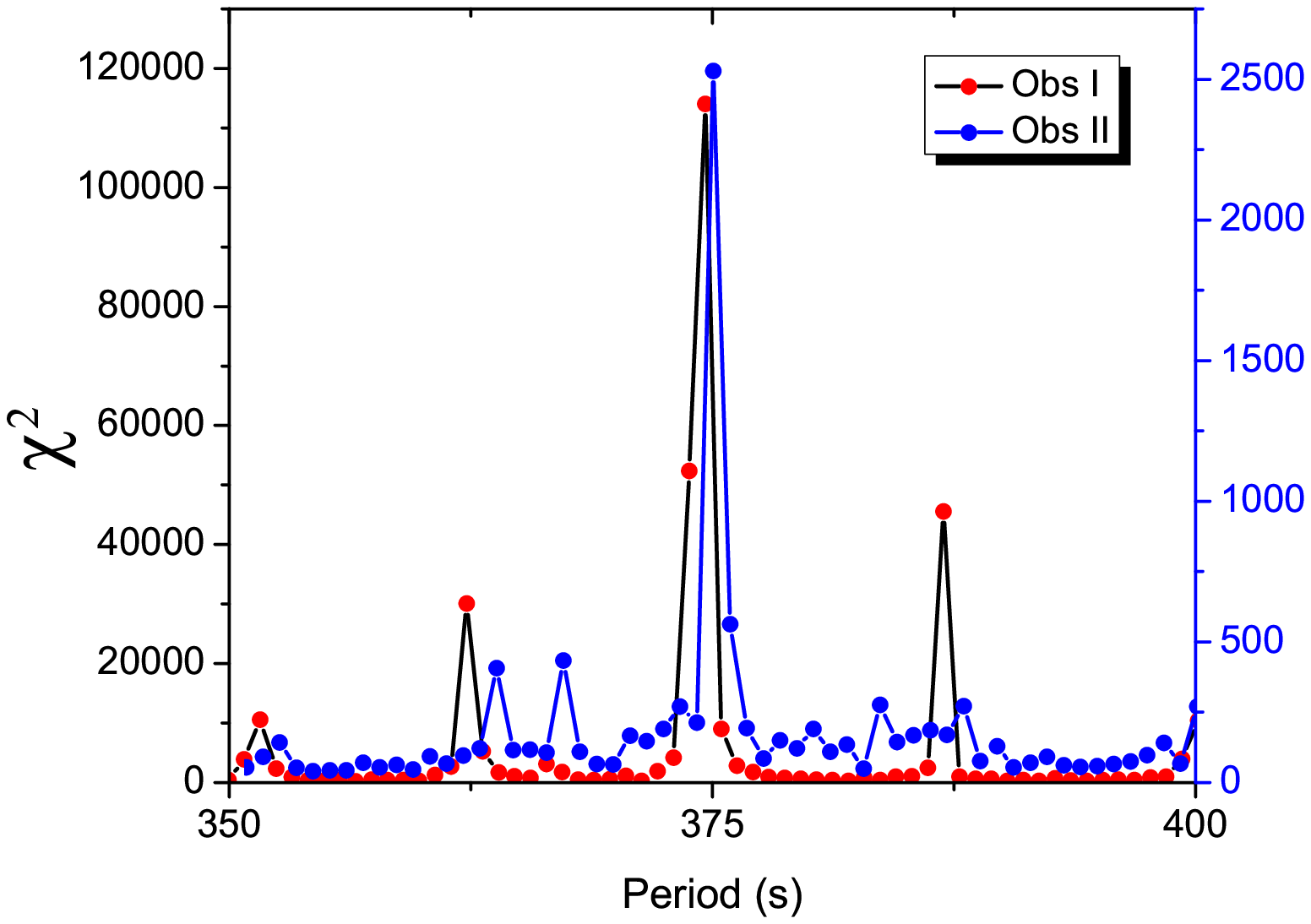}
\end{center}
\caption{Periodogram of the source corresponding to NuSTAR observations.}
\label{1}
\end{figure}

\begin{figure}

\begin{center}
\includegraphics[angle=0,scale=0.7]{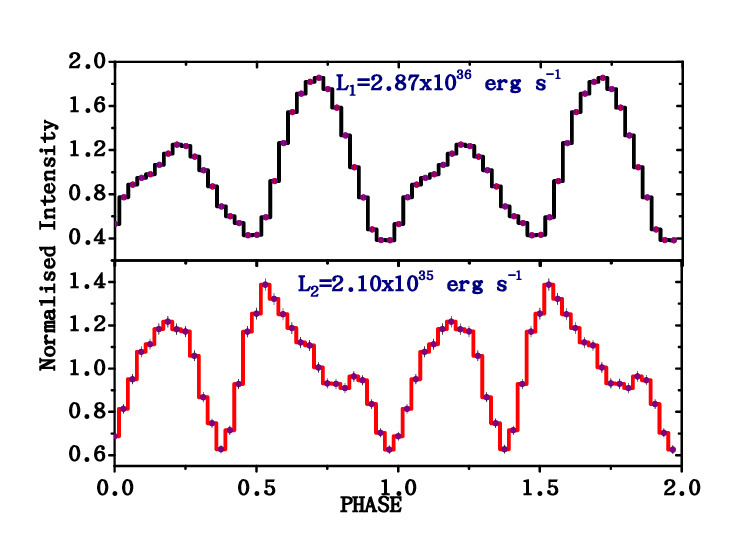}
\end{center}
\caption{Pulse profiles of IGR J06074+2205 for Obs I \textit{(upper panel)} and Obs II \textit{(lower panel)} in (3-79) keV energy range highlighting the respective luminosity measurements.}
\label{3}
\end{figure}

\begin{figure}
\begin{minipage}{0.3\textwidth}

\includegraphics[angle=0,scale=0.75]{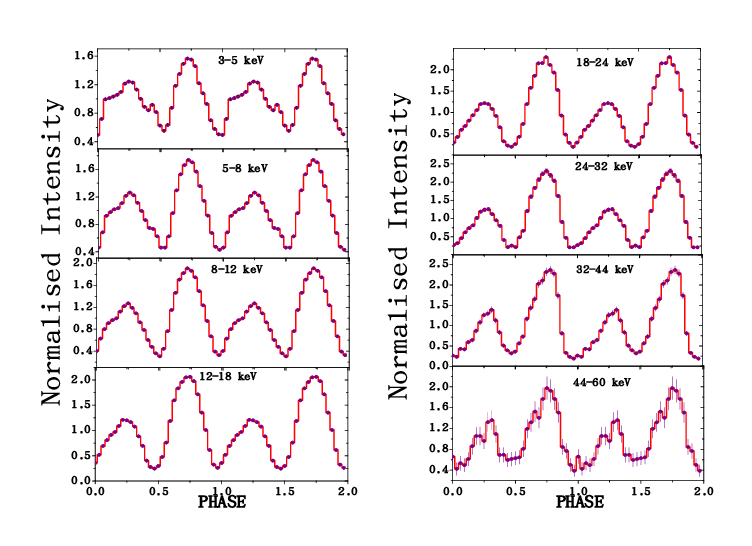}
\end{minipage}
\hspace{0.15\linewidth}
\begin{minipage}{0.3\textwidth}
\includegraphics[angle=0,scale=0.75]{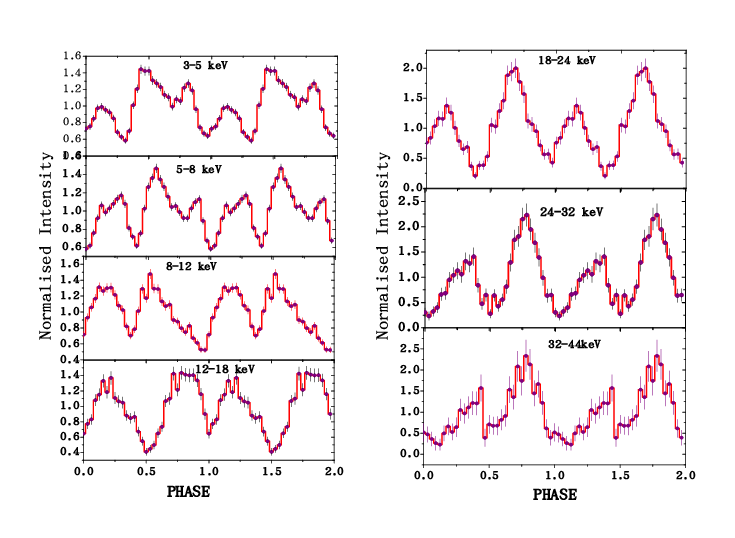}
\end{minipage}
\hspace{0.15\linewidth}

\caption{Energy-resolved pulse profiles of IGR J06074+2205 -- Obs I \textit{(top)} and Obs II \textit{(bottom)}}
\label{4}
\end{figure}

 The investigation into the pulsation behavior of the source involved a thorough analysis, employing various techniques to unveil the intricate characteristics of its periodic signals. Initially, a Fast Fourier Transform (FFT) is applied to the 0.1 s binned light curve, serving as an effective tool to approximate the pulse period of the source. We performed the epoch-folding technique \citep{16,17}  based on  $\chi^{2}$ maximization for precise determination of the pulse period. The best period for the source was estimated to be 374.626(9) and 374.661(8) s for observation I and II respectively. The measurement uncertainties were approximated using the method proposed by \cite{Boldin}. This involved generating 1000 simulated light curves, allowing for a robust estimation of the uncertainty associated with the derived pulse period. \textbf{Data are not corrected for the effects of the binary system due to the fact that the orbital solutions of this source have not yet been thoroughly investigated.} The tools EFSEARCH and EFOLD are used  for the determination of the optimum pulse period and generation of the folded pulse profiles. The Periodogram of the source, for NuSTAR  observations is presented in Figure \ref{1}. Incorporating historical spin period data from the Fermi/Gamma-ray Burst Monitor (GBM) \citep{Malacariaa} added a temporal dimension to the analysis. The observed spin-down rate of the source, quantified at 0.0202(2) s day$^{-1}$, which provides crucial information about the spin period evolution of the pulsar.
 \begin{figure}

\begin{center}
\includegraphics[angle=0,scale=0.35]{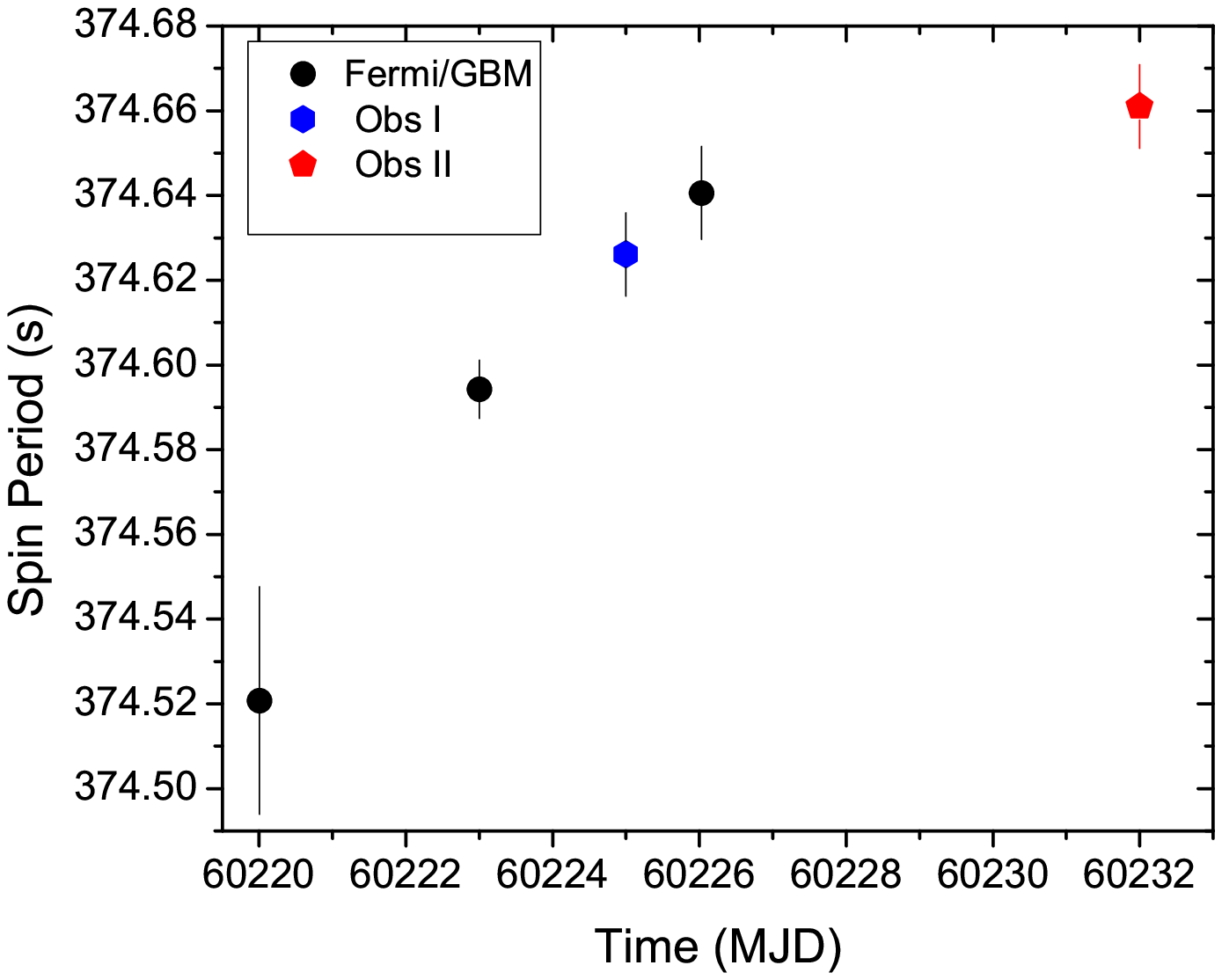}
\end{center}
\caption{Spin period evolution of IGR J06074+2205 from recent Fermi-GBM and NuSTAR observations.}
\label{2}
\end{figure}
The analysis is extended by examining the energy dependence of the pulse profiles. By resolving the light curve in the (3-79) keV energy range into distinct energy bands, energy-resolved pulse profiles were generated. These profiles revealed intriguing variations in the pulsation characteristics across different energy levels, offering valuable insights into the physical processes underlying the emission.
Therefore, the combined application of Fourier techniques, epoch-folding, and energy-resolved profiles have enabled a comprehensive understanding of the temporal features of the source. The incorporation of historical spin period data has enhanced our knowledge of the spin period evolution of the source. Figure \ref{2}  represents the evolution of the spin period over the observed period. Additionally, the pulse profiles presented in Figure \ref{3} and Figure \ref{4} provide a detailed view of how the pulsation characteristics changed across different energy bands.
   
 The pulse fraction (PF) is a measure of the relative amplitude of the evolving pulse profiles which is defined as 
  $PF=\;\frac{P_{max}-P_{min}}{P_{max}+P_{min}}$, where $P_{max}$ and $P_{min}$ represent the maximum and minimum intensities of the emerging pulse profile, respectively. Observations of various X-ray pulsars, as documented by \cite{37}, reveal a notable trend in the pulse fraction with respect to energy. Specifically, the pulse fraction of the source tends to exhibit an overall increasing trend with increasing energy levels, which is related to the energy-dependent modulation of the X-ray emission, with higher-energy photons contributing more significantly to the observed pulsations.

\begin{figure}

\begin{center}
\includegraphics[angle=0,scale=0.4]{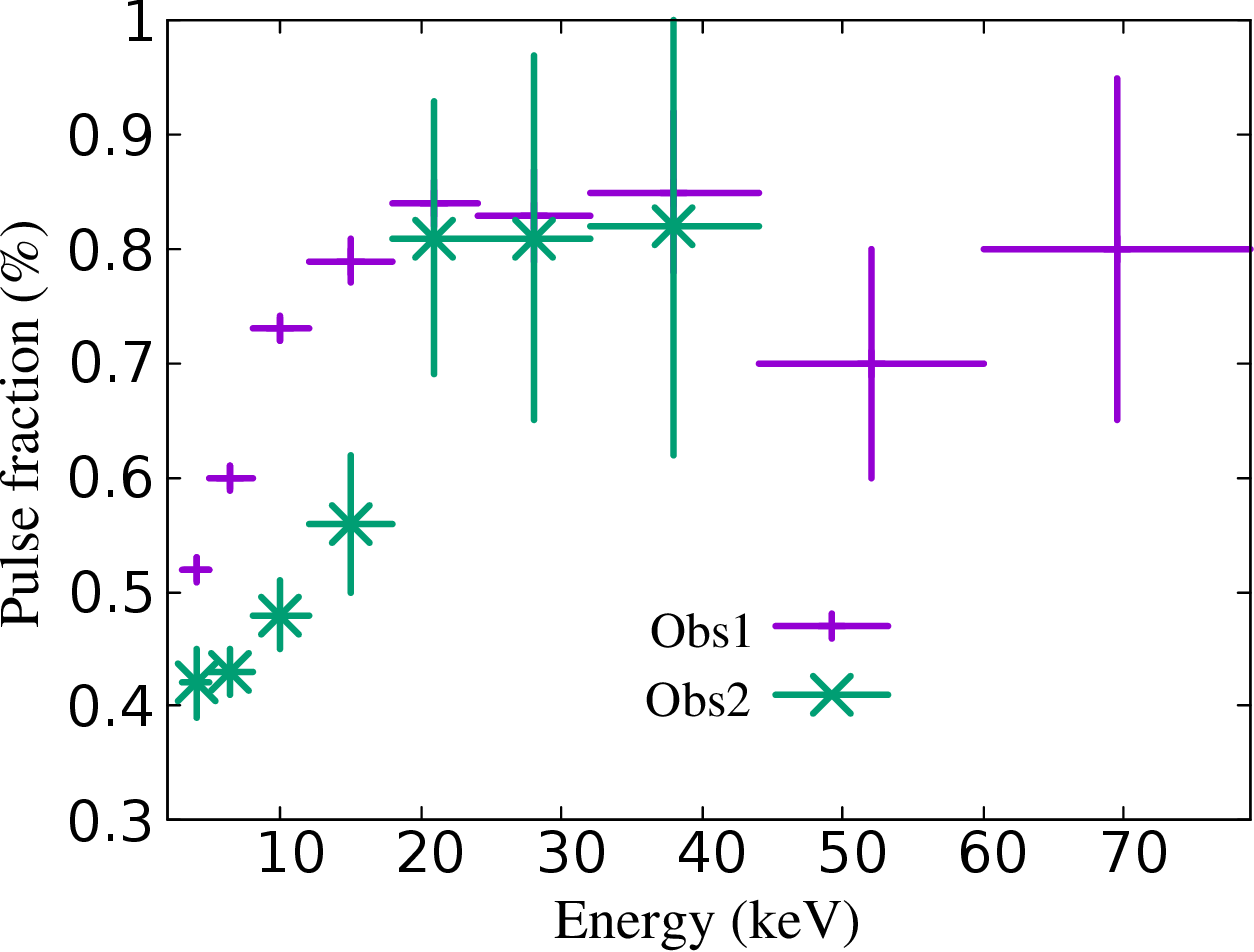}
\end{center}
\caption{Variation of pulse fraction with energy corresponding to NuSTAR observations of IGR J06074+2205.}
\label{5}
\end{figure}

 Notably, there is a tendency for the pulse fraction to drop at  an energy level above  (45-50) keV as evident from Figure \ref{5}. The substantial errors in the PF of the higher energy bins make this drop insignificant. It is currently difficult to definitively state whether the presence of the absorption feature leads to the drop in PF.
 
However, there have been multiple indications of a distinct drop in the pulse fraction specifically around the energy associated with the cyclotron line. The phenomenon has been reported in the literature, e.g., \cite{TSY, 39, 37, tsy, lut}. The cyclotron line, representing a characteristic absorption feature in the X-ray spectrum, is related to a modulation in the observed pulsations. This modulation manifests as a reduction in the pulse fraction, providing valuable indirect evidence for the presence of the cyclotron line and it highlights the impact on the pulsar's emission properties.


\section{Spectral Analysis}
The NuSTAR X-ray spectral analysis of IGR J06074+2205 provides valuable insights into the spectral properties of the source across the broad energy range of 3.0-79.0 keV. Several continuum models, such as CUTOFFPL, HIGHECUT, COMPTT, etc., were employed in the fitting process. Notably, the most robust fit was done using the  CUTOFFPL model, emphasizing its suitability in describing the observed spectral features. The NuSTAR data from both FPMA and FPMB modules are effectively utilized in the analysis. To ensure reliable and statistically significant results, the data are  grouped using GRPPHA, with a minimum of 25 counts per spectral bin. The relative normalization factors between the two NuSTAR modules were carefully maintained by freezing the constant factor corresponding to FPMA as unity, while allowing FPMB's constant factor to vary. This methodology ensured comparable count rates between the two instruments, enhancing the reliability of the spectral analysis.

\begin{figure}
\begin{minipage}{0.3\textwidth}

\includegraphics[angle=0,scale=0.33]{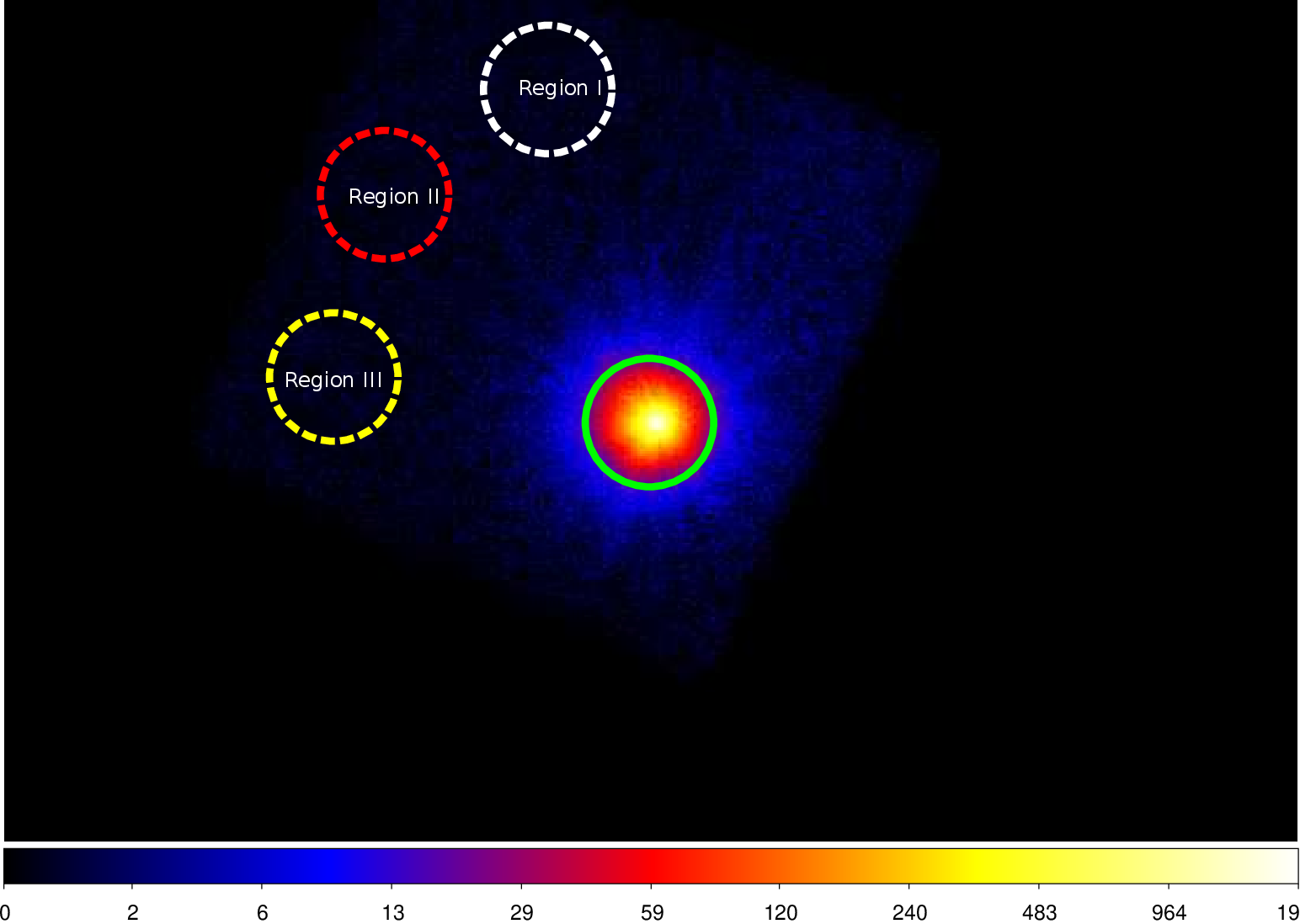}
\end{minipage}
\hspace{0.15\linewidth}
\begin{minipage}{0.3\textwidth}
\includegraphics[angle=0,scale=0.33]{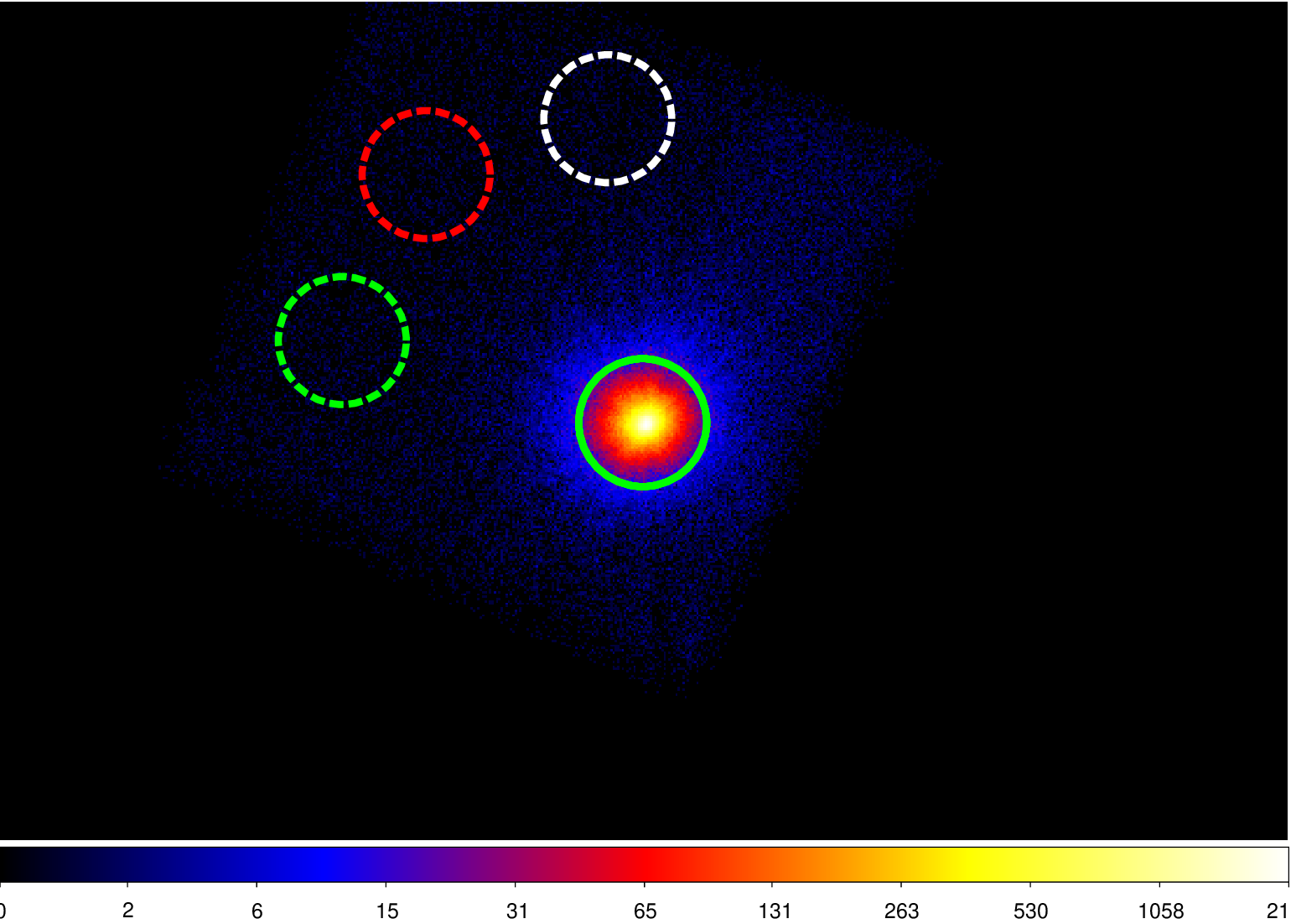}
\end{minipage}
\hspace{0.15\linewidth}

\caption{Image of the source obtained with FPMA (\textit {upper panel}) and FPMB (\textit{lower panel}), scaled logarithmically with three different choices of background regions : Region I (white), Region II (red), and Region III (green), each of 90 arcsec radius for Obs I.}
\label{6}
\end{figure}

The absorption column density was modeled using the TBABS component with abundance from \cite{Wilms}, and the cross-section was determined using \textit{vern} \citep{Verner}. The initial fitting utilized the continuum model CONST*TBABS*CUTOFFPL. However, distinct residuals are observed in the X-ray continuum, which encouraged further investigation. An iron emission-line at $\sim$ 6.28 keV with an equivalent width of $\sim$14 eV was observed in the spectra which was fitted using a GAUSSIAN component.  We performed \textit{F-test} to determine if the model with the line is preferable.  The low F-test probability of $10^{-5}$ suggests that the emission feature is significant. Such a narrow iron line suggests that the emission originates from a small, localized area, associated with the accretion disk. It may also indicate the presence of an optically thick, compact accretion disk that is capable of effective X-ray reflection \citep{Jaisawal}. To account for the negative residuals, a Gaussian absorption component (GABS) was introduced. This adjustment effectively addressed the observed negative residuals, providing a more accurate representation of the source's X-ray continuum. The broad absorption feature observed in the X-ray continuum may be interpreted as a cyclotron line, a characteristic absorption feature observed in the X-ray spectra of certain astrophysical sources. The analysis pointed out that the cyclotron line is characterized by a centroid energy of $\sim$55 keV, with an equivalent width of about 13.5 keV. The observed line energy  is marginally higher than that reported by \cite{Roy}. The inclusion of this feature significantly improved the spectral fit, underlining its importance accurately describing the observed data. In order to test the significance of the characteristic absorption
line, we simulated the spectra of the NuSTAR observation using the XSPEC script \textit{simftest}. The best-fit $\chi^{2}$ values were determined by fitting spectral models with and without the absorption feature in  $10^{4}$ simulations. The probability of false-detection was found to be less than $\ 10^{-5}$. The flux in the 3.0-79.0 keV energy range is estimated to be $\sim1.19\;\times\;10^{-9} erg\;cm^{-2}\;s^{-1}$, which corresponds to an X-ray luminosity of $\sim2.87\;\times\;10^{36} erg\;s^{-1}$ considering a source distance of 4.5 kpc \citep{Reigg}.
Additionally, to verify the extent to which the spectral parameters are affected by the selection of the background region at  different positions, we performed spectral analysis for three background regions (Figure \ref{6}). Table \ref{c} illustrates that there is no appreciable effect of background selection on the spectral data.

Similarly, the best spectral fit for obs II was obtained using the CUTOFFPL model. We neglected the spectra above 50 keV  due to background contamination.  Distinct negative residuals were observed in the X-ray continuum at $\sim$ 10 keV, which was fitted by introducing a GABS component. This led to a significant improvement of the fit statistics from (1019/733) to (760/731). Given that NuSTAR responses show larger uncertainty around the tungsten edge at $\sim$10 keV, the physical significance of this feature remains unexplained and could potentially have an instrumental origin \citep{Madsen}. This peculiar spectral feature at $\sim$ 10 keV has been observed in various sources and was investigated in a recent study by \cite{Manikantan}. The recent study by \cite{Roy} did not discuss or make reference to this particular characteristic. Instead, they reported a soft excess in the spectrum. The source flux was found to drop significantly to $\sim8.67\;\times\;10^{-11} erg\;cm^{-2}\;s^{-1}$ during this observation. 

\begin{figure}
\begin{minipage}{0.3\textwidth}

\includegraphics[angle=0,scale=0.35]{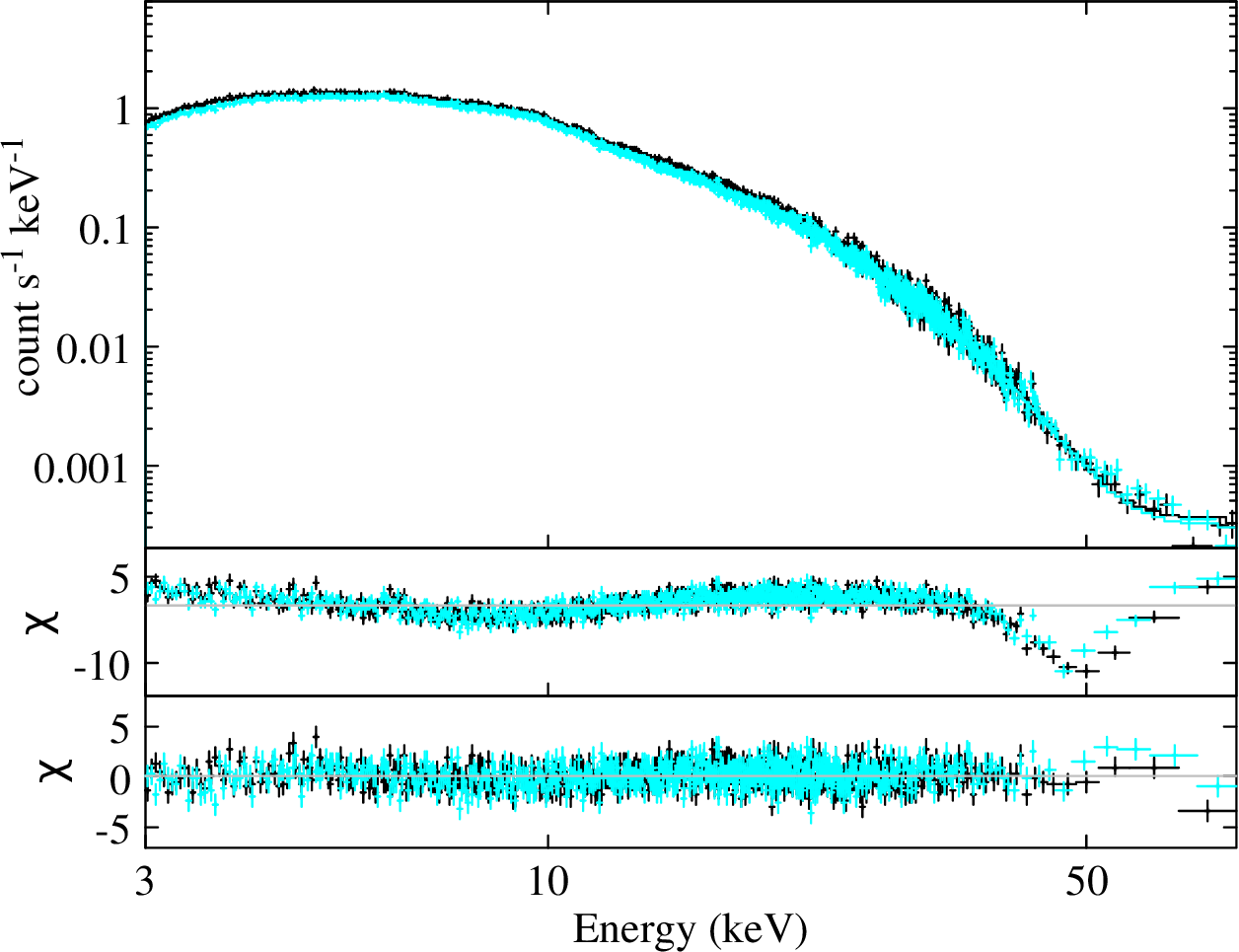}
\end{minipage}
\hspace{0.15\linewidth}
\begin{minipage}{0.3\textwidth}
\includegraphics[angle=0,scale=0.36]{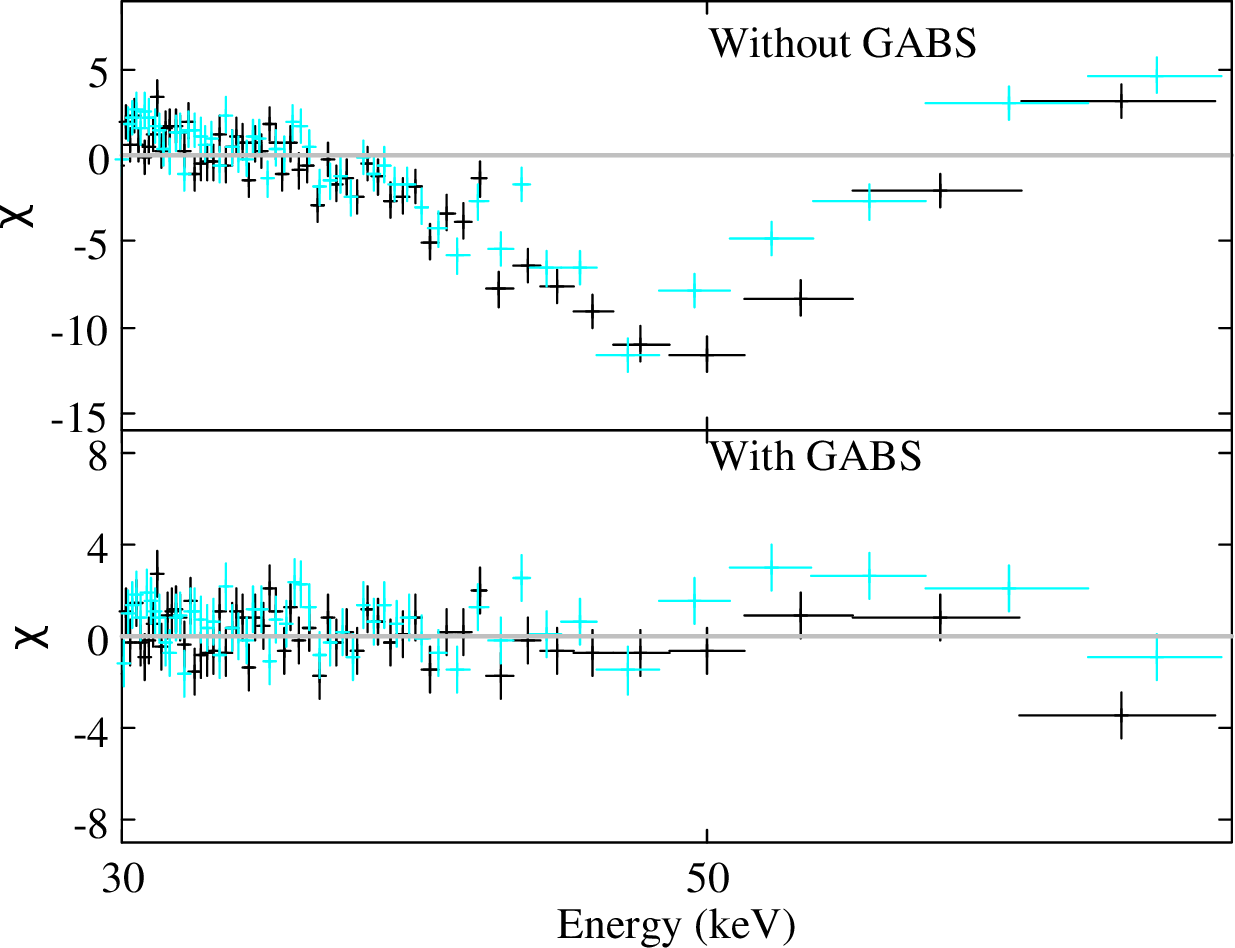}
\end{minipage}
\hspace{0.15\linewidth}

\caption{The best fit spectra of the source (Obs I) is shown in the upper panel (black for FPMA and cyan for FPMB.) The middle and bottom panels depict residuals before and after incorporating the absorption component. (\textit{top}).
Zoomed section in 30 to 79 keV range rebinned  for plotting purposes (\textit{bottom}).}
\label{7}
\end{figure}

\begin{figure}

\begin{center}
\includegraphics[angle=270,scale=0.37]{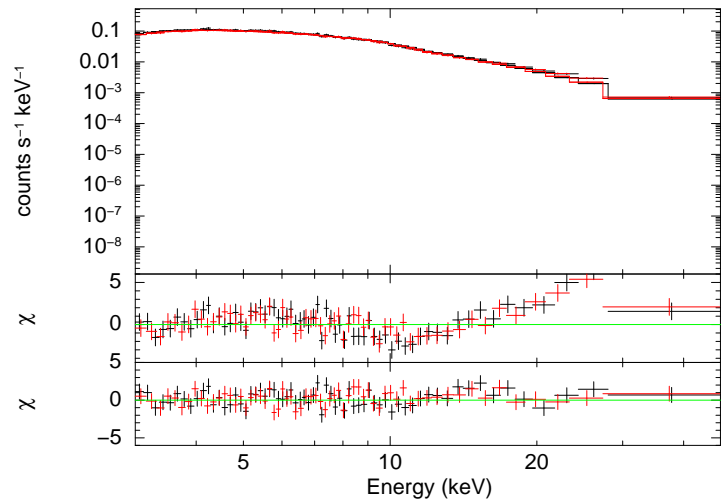}
\end{center}
\caption{The best fit spectra of the source (Obs II) is shown in the upper panel (black for FPMA and red for FPMB). The middle and bottom panels depict residuals before and after incorporating the GABS component.} 
\label{8}
\end{figure}

The best-fit spectral parameters obtained from this comprehensive modelling approach are summarized in Table \ref{b}. The NuSTAR spectra of the source are presented in Figure \ref{7} (Obs I) and Figure \ref{8} (Obs II), providing a visual representation of the spectral characteristics and the effectiveness of the adopted modelling approach here.
\begin{table}
\begin{center}
\begin{tabular}{ccccc}
\hline										
 Spectral Parameters &  & Obs I& &Obs II  \\
\hline							
$C_{FPMA}$		&	&	1.0(fixed)	&	&	1.0 (fixed)\\
$C_{FPMB}$		&	&	$0.997^{+0.001}_{-0.001}$&	&	$1.004^{+0.001}_{-0.001}$\\
$nH\;(\times 10^{22})\;(cm^{-2})$		&	&$0.49^{+0.11}_{-0.13}$		& &$0.58^{+0.11}_{-0.19}$		\\
$\Gamma$		&	& $1.09^{+0.05}_{-0.04}$&	& $0.99^{+0.15}_{-0.11}$\\
$\Gamma_{norm}(\times10^{-2})$		&	& $5.69^{+0.18}_{-0.15}$&	& $1.60^{+0.11}_{-0.12}$\\
$E_{cut}$ (keV)		&	&$38.36^{+0.001}_{-0.001}$	&	&$16.81^{+1.03}_{-1.07}$	\\

$E_{Fe}$(keV)	&	&	$6.28^{+0.09}_{-0.11}$&	&--\\
$\sigma_{Fe}$ (keV) &	&	$0.10^{+0.04}_{-0.07}$&	&--\\
$norm_{Fe}(\times10^{-4})$  &	&	$1.66^{+0.21}_{-0.21}$&	&--\\
$EW_{Fe}$ (eV) &	&	$14^{+4}_{-4}$&	&--\\
$E_{gabs}$(keV)	&	&	--&	&$10.27^{+1.02}_{-0.95}$\\
$\sigma_{gabs}$ (keV)&	&--	&	&$3.33^{+0.88}_{-1.02}$\\
$\tau_{gabs}$		&	&--		& &$1.02^{+0.17}_{-0.21}$	\\
$E_{cyc}$(keV)		&	&	$54.61^{+1.15}_{-0.91}$&	&--\\
$\sigma_{cyc}$ (keV)	&	&	$13.53^{+1.17}_{-1.21}$	&	&--\\

$\tau_{cyc}$		&	&	$1.24^{+0.14}_{-0.11}$	&	&--\\
$\chi^{2} /d.o.f.$		&	& 1804/1754		&   &760/731	\\
$\chi^{2}_{\nu}$		&	&	1.03	&	& 	1.04\\

 \hline
 \end{tabular}
 \caption{The best fit parameters of IGR J06074+2205 in 3.0-79 keV energy range. The fit statistics $\chi_{\nu}^{2}$  represents reduced $\chi^{2}$ ($\chi^{2}$ per degrees of freedom). Errors quoted for each parameter are within 90\% confidence interval. The parameters are -
  nH : hydrogen column density, $\Gamma$ : photon-index, $E_{cut}$ : cutoff energy,  $E_{fold}$ : folding energy. $E_{Fe}$ , $EW_{Fe}$ :  line energy and equivalent width of the iron emission line. $E_{cyc}$, $\sigma_{cyc}$, and $\tau_{cyc}$ : centroid energy,  width  and optical depth of the cyclotron line. Parameters for the '10 keV feature' in Observation II are represented using the 'gabs' subscript.}
  \label{b}
  \end{center}
  
 \end{table}

 \begin{table*}
\begin{center}
\begin{tabular}{ccccc}
\hline	
Background &  $REG_{I}$ & $REG_{II}$ &$REG_{III}$ \\
\hline							
$C_{FPMA}$		&	1.0(fixed)	&	1.0 (fixed)&1.0(fixed)\\
$C_{FPMB}$		&	$0.997^{+0.001}_{-0.001}$&$1.002^{+0.001}_{-0.001}$&$1.003^{+0.001}_{-0.001}$\\
$nH\;(\times 10^{22})\;(cm^{-2})$		&	$0.49^{+0.11}_{-0.13}$		& $0.38^{+0.09}_{-0.12}$ & $0.44^{+0.10}_{-0.16}$	\\
$\Gamma$		&	 $1.09^{+0.05}_{-0.04}$&	 $1.04^{+0.11}_{-0.14}$ &	 $1.07^{+0.15}_{-0.11}$\\
$E_{cut}$ (keV)		&	$38.36^{+0.001}_{-0.001}$	&	$40.73^{+1.05}_{-1.11}$ &	$41.81^{+1.04}_{-1.09}$	\\

$E_{Fe}$(keV)		&	$6.28^{+0.09}_{-0.11}$&	$6.27^{+0.08}_{-0.14}$&	$6.27^{+0.13}_{-0.10}$\\
$\sigma_{Fe}$ (eV)	&	$14^{+4}_{-4}$&	$14^{+2}_{-3}$ &	$13^{+4}_{-4}$ \\
$E_{cyc}$(keV)		&	$54.61^{+1.15}_{-0.91}$&$54.52^{+1.11}_{-1.18}$	&$54.87^{+1.09}_{-0.91}$\\
$\sigma_{cyc}$ (keV)	&	$13.53^{+1.17}_{-1.21}$	&	$13.48^{+1.11}_{-1.17}$	&$13.72^{+1.21}_{-1.14}$	\\
$\tau_{cyc}$		&	$1.24^{+0.14}_{-0.11}$	&	$1.22^{+0.11}_{-0.12}$&	$1.26^{+0.16}_{-0.14}$\\
$\chi^{2} /d.o.f.$		&	 1804/1754		&   1879/1778	&   1885/1778\\
$\chi^{2}_{\nu}$		&	1.03	&	 	1.06&	 	1.06\\

 \hline
 \end{tabular}
 \caption{The best fit parameters of IGR J06074+2205 (Obs I) in 3.0-79 keV energy range for three different background regions.  Errors quoted for each parameter are within 90\% confidence interval.}
 \label{c}
  \end{center}
  
 \end{table*}

\section{Discussion and Conclusion}

The study of the BeXRB source IGR J06074+2205, based on recent NuSTAR data, provides a comprehensive exploration of its broadband spectro-temporal characteristics. The coherent X-ray pulsation of the source is successfully identified through timing analysis of the photon events, indicating  a pulse period of $\sim$ 374.62 s. Furthermore, the analysis of the source's spin period history indicates a discernible spin-down trend at a rate of 0.0202(2) s day$^{-1}$. The  results obtained here collectively contribute to a detailed understanding of the rotational and  the temporal behaviour of the source.

A significant and interesting aspect of the study is the observed energy dependence in the pulse patterns of IGR J06074+2205. The energy-resolved profiles exhibit a dual-peaked pattern, with a noteworthy decrease in the amplitude of the secondary peak and a proportionate increase in the amplitude of the primary peak that increases with energy. This evolving morphology in the pulse profiles indicates a dynamic and changing accretion state of the pulsar, which provides a  valuable insight into the underlying astrophysical processes that are going on.

The fraction of photons contributing to the modulated section of the flux, commonly referred to as pulse fraction, emerges as a key parameter in understanding the source characteristics. The observed distinct absorption feature in the X-ray continuum aligns with the energy-dependent variation of PF. Specifically, PF demonstrates an overall increasing tendency with energy. This strange behaviour is further elucidated using a toy model proposed by \cite{landt}, which indicates that the increasing trend of PF with energy may be attributed to the X-ray producing zone that is associated with a more compact nature at higher energies. This compactness enhances the pulsed emission of the neutron star, contributing to the observed energy-dependent variations in PF. The source spectrum exhibits a significant CRSF, providing further insight into the source's emission properties. It is noteworthy that uncertainties in the PF measurements are relatively higher at higher energies due to limited statistics associated with the pulsed emission at these energy levels.

 The  absorbed flux of the BeXRB source IGR J06074+2205 in the 3-79 keV energy range is estimated to be valuable insights into its energetics. The estimated absorbed flux, approximately $\sim1.19\;\times\;10^{-9} erg\;cm^{-2}\;s^{-1}$, when considered with a source's distance of 4.5 kpc, it translates to an X-ray luminosity of $\sim 2.87 \times 10^{36} \;{erg}\;{s}^{-1}$. This luminosity is a critical parameter in understanding the energetics of the source and provides a basis for further analysis.

CRSFs are identified as absorption line features in the hard X-ray range \citep{Staubert}. The centroid energies of CRSFs have been found to vary in the range (5-100) keV. For the observed centroid energy of the cyclotron line, estimated to be around $\sim 55$ keV, a magnetic field strength of  $\sim 4.74 \times 10^{12}$ G is predicted. Such high values of cyclotron line energy have been reported in several sources, including GX 304-1, V 0332+53, A 0535+26, GRO 1008-57, 1A 1118-616, etc,. \citep{Staubert}. The critical luminosity ($L_{crit}$), as proposed by \cite{Becker}, marks the transition between different accretion states. Based on the maximum value for the CRSF centroid energy and standard neutron star parameters, $L_{crit}$ may be expressed as: 

 $L_{crit}$ = 1.5 $\times10^{37}\;B_{12}^{16/15}$ erg\; $s^{-1}$,  
where $B_{12}$ represents the surface magnetic
field strength expressed in units of $10^{12} G$. We estimated $L_{crit}$ to be $\sim 7.89 \times 10^{37} erg\; s^{-1}$. It is noteworthy that the estimated X-ray luminosity of the source is lower than $L_{crit}$.  This observation suggests that the source may be accreting in the subcritical accretion regime. In this regime, the beaming pattern is typically dominated by a pencil-beamed configuration, where accreted material penetrates the surface of the neutron star through processes like nuclear collisions with atmospheric protons or Coulomb collisions with thermal electrons \citep{Harding}. In such a scenario, the emission arises from the top of the accretion column \citep{Burnard}. The luminosity measurements in observation II reveal a further decrease to $\sim 2.10 \times 10^{35} erg\; s^{-1}$. According to \cite{Mushtukov}, accretion on the surface of neutron stars occur at near free fall velocities, leading to the formation of accretion mounds or hot spots near the neutron star's polar caps. These hot spots dominate the radiation observed from the system.

Looking ahead, future investigations may focus on exploring variabilities associated with the spectral parameters, including characteristics of the cyclotron line. With the availability of additional observational data for the source, a more detailed understanding of its accretion  and emission mechanisms may be  determined. The subcritical accretion regime identified in this study sets the stage for further exploration of the intricate astrophysical processes governing the interesting observation of X-ray emission from IGR J06074+2205.

\section{DATA AVAILABILITY}
The observational data employed in this research work can be accessed from the HEASARC data archive and is publicly available.

\section{Acknowledgements}
The research work has been performed using publicly available data of the source provided by NASA HEASARC data archive. We would like to thank ICARD, University of North Bengal, Department Of Physics, for providing research facilities to carry out this work. We would like to thank the anonymous reviewer for the valuable time and  insightful remarks  to improve the manuscript.


\end{document}